\documentclass[a4paper,12pt]{article}
\usepackage{amsmath,amssymb}
\usepackage{epsfig}
\begin{document}

\begin{flushright}
JINR-P2-82-900\\
\end{flushright}
\vspace{2.0cm}

\begin{center}
{\large\bf  Anomalous dimensions of quark masses in the three-loop
approximation \\}
\vspace{1.5cm}
{\large O.V. Tarasov}\\
\vspace{0.5cm}
{\em Joint Institute for Nuclear Research, Dubna, Russia}\\
\end{center}
\vspace{3cm}
\begin{abstract}
The results of calculation of the three-loop radiative correction to
the renormalization constant of fermion masses for non-abelian gauge theory
interacting with fermions are presented.
Dimensional regularization and the t'Hooft-Veltman  minimal subtraction scheme   
 are used. The method of calculation is described in detail.
The renormalization group function $\gamma_m$ determining the behavior 
of the effective mass of fermions is 
presented. The  anomalous dimensions of fermions for QED and QCD up to 
three loops are given.
All calculations were performed on a computer with the help of
the  SCHOOONSCHIP system for analytical manipulations.\\

\end{abstract}
\newpage
\vspace{0.3cm}
\vspace{1cm}
The 
effects determined by the nonzero quark and lepton masses
in quantum chromodynamics, the  standard $SU(2)\times U(1)$ model,
usual and supersymmetric grand unified theories are now attracting
 now more and more attention. Considerable  efforts were devoted 
 to determining the mass of the Higgs
scalar particle from the LEP experiments. The threshold effects \cite{Shirkov:1981mb},
the light quark masses \cite{Kataev:1982xu} and the value of the ratio $m_b/m_\tau$ in the
frame of the grand unified theories \cite{Nanopoulos:1981mv}
are usually investigated by the renormalization group (RG)
method \cite{Vladimirov:1979ib}. Application of the RG allows one to sum 
the leading momentum-dependent logarithms to all orders in terms of the limited
number of the RG parameters (beta functions and anomalous dimensions) which
are calculated in the framework  of perturbation theory.

The most characteristic feature of the mass account is large contributions
of the higher orders in  perturbation theory. For example, the two-loop
corrections to  the ratio $m_b/m_\tau$ \cite{Nanopoulos:1981mv}
are 20\% from the leading contribution. In the $SU(5)$ grand unified 
theory from two-loop analysis it was concluded \cite{Nanopoulos:1981mv} 
that the number of flavors cannot exceed six. The one - loop approximation gives a 
less rigid restriction- the number of flavors may be less than eight. 
The next approximations -  the three
loop and higher, can also change this restriction. Higher corrections to the
renormalization group functions depend more strongly  on the number
of quarks.

We consider the non-Abelian gauge theory with fermions belonging to the
representation $R$ of the gauge group G:
$$
L=-\frac{1}{4}G_{\mu \nu}^aG_{\mu \nu}^a-\frac1{2 \alpha}\left(\partial
_\mu A^a_\mu\right)^2-\partial_\mu \overline{\eta}^a \partial _ \mu \eta^a+\\
gf^{abc} \overline{\eta}^a A^b_\mu \partial_ \mu \eta^c+
i\sum_{l=1}^{f} \overline{\psi}^l_i \hat{D} \psi^l_i
-\sum_{l=1}^{f} m_l \overline{\psi}^l_i \psi^l_i,
$$
\begin{eqnarray*}
& & G^a_{\mu \nu} = \partial _\mu A^a_{\nu}-\partial _\nu A^a_{\mu}+
gf^{abc} A_{\mu}^b A_{\nu}^c, \cr
& & D_{\mu} \psi^l_i=\partial_ \mu \psi^l_i -igR^a_{ij} \psi_j^l A^a_\mu.
\end{eqnarray*}

Here $\eta^a$ is the  ghost field, $\alpha$ is the gauge parameter
and $f^{abc}$ are totally antisymmetric structure constants of the
gauge group $G$. The indices of the fermion field $\psi_i^m$ specify color ($i$)
and flavor ($m$), respectively. The matrices $R^a$ obey the following
relations:
$$
\left[R^a,R^b\right]_-=if^{abc}R^c,~~~~ f^{acd}f^{bcd}=C\delta^{ab},
$$
$$
R^aR^a=C_FI,~~~~~tr\left(R^aR^b\right)=t\delta^{ab}.
$$

In particular, the values of group invariants $C$, $C_F$ and $T$ in the
fundamental (quark) representation of $SU(N)$ are:
\begin{eqnarray*}
C=N,~~~~~~C_F=\frac{N^2-1}{2N},~~~~~~t=\frac{1}{2}.
\end{eqnarray*}

In this paper we adopt the renormalization prescription 
by 't Hooft  \cite{'tHooft:1973mm}
the so-called "minimal subtraction scheme" (MS), which by definition subtracts
only pole parts in  $\varepsilon$ from a given diagram. The renormalization
constant $Z_{\Gamma}$ relating the dimensionally regularized 1PI Green
function with the renormalized one
\begin{eqnarray*}
\Gamma_R\left(\frac{Q^2}{\mu^2},h,m,\alpha\right)=\lim_{\varepsilon \rightarrow 0}
Z_\Gamma \left(\frac{1}{\varepsilon},h,\alpha\right)
\Gamma \left(Q^2,h_B,m_B,\alpha_B,\varepsilon \right),
\end{eqnarray*}
looks in this scheme like
\begin{eqnarray*}
Z_\Gamma \left(\frac{1}{\varepsilon},h,\alpha\right)=
1+\sum_{\nu=1}^{\infty}c_\Gamma
^{(\nu)}(h,\alpha)\frac{1}{\varepsilon^{\nu}},
\end{eqnarray*}
with $\varepsilon=(4-n)/2$, $n$  being the space-time dimension,
$h=g^2/(4\pi)^2$ and
$\mu$ is the renormalization parameter with the dimension of mass.
The relationship between the bare charge $h_B^2$ and the
renormalized one includes the product of appropriate
$Z$'s.  The most convenient choice of $Z$'s is as follows:
\begin{eqnarray*}
h_B=(\mu^2)^\varepsilon h\widetilde Z_1^2Z_3^{-1} \widetilde Z_3^{-2}=
(\mu^2)^\varepsilon
\left[h+\sum_{\nu=1}^{\infty}a^{(\nu)}(h)\frac{1}{\varepsilon^{\nu}}\right].
\end{eqnarray*}
Here $\widetilde{Z}_1$ is the renormalization constant of the ghost-ghost-gluon
vertex, $Z_3$ and $\widetilde{Z}_3$ being those of the inverse
gluon and ghost propagators, respectively. Bare parameters $m_B$ and
$\alpha_B$ are connected with the renormalized ones as:
\begin{eqnarray}
& &m_B=Z_mm=m\left[1+\sum_{\nu=1}^{\infty}b^{(\nu)}(h)\frac{1}
{\varepsilon^{\nu}}\right],
\nonumber
\\
& &\alpha_B=Z_3\alpha=\alpha\left[1+\sum_{\nu=1}^{\infty}d^{(\nu)}(h,\alpha)
\frac{1}{\varepsilon^\nu}\right].
\label{massaB}
\end{eqnarray}

In Ref.\cite{Caswell:1974cj} it was shown that $a^{(\nu)}(h)$ 
and $b^{(\nu)}(h)$ in the $\overline{MS}$ renormalization
 scheme are gauge independent. This gives us a 
possibility to use the simplest gauge for calculation. We choose 
to work in the Feynman gauge $\alpha =1$ throughout this paper.

The Green function $\Gamma_R \left(\frac{Q^2}{\mu^2},h^2,m,\alpha \right)$
in the $\overline{MS}$ scheme
satisfies the renormalization group equation \cite{'tHooft:1973mm},
\cite{Collins:1973yy}:
\begin{eqnarray}
&&\left\{ Q^2 {\partial \over \partial Q^2}-\beta(h)
{\partial \over \partial h}
+\left[1-\gamma_m(h) \right]m{\partial \over \partial m}
 \right.
\nonumber \\
&&\left.~~~~~~~~~~~~~~~~~~~~~~~-  \delta (h,\alpha)
\alpha {\partial \over \partial \alpha}-\gamma_{\Gamma} (h,\alpha)
 \right\} \Gamma_R \left( \frac{Q^2}{\mu^2},h,m,\alpha \right)=0. 
\label{RGequatio}
\end{eqnarray}
Here
\begin{eqnarray*}
& & \beta (h)=\left. \mu^2 {\partial h \over \partial \mu^2}
                              \right|_{h_B,m_B,\alpha_B-fixed}, \cr
&& \\			      
& & \gamma_m (h)= \left. \mu^2 {\partial \ln m \over \partial \mu^2}
                               \right|_{h_B,m_B,\alpha_B-fixed},\cr
& & \delta (h,\alpha)=\left. \mu^2 {\partial \ln \alpha \over \partial \mu^2}
                               \right|_{h_B,m_B,\alpha_B-fixed},\cr
& & \gamma_{\Gamma} (h,\alpha)=-\left. \mu^2 {\partial \ln Z_{\Gamma} \over
     \partial \mu^2}\right|_{h_B,m_B,\alpha_B-fixed}.
\end{eqnarray*}

The functions $\beta(h),\gamma_m(h),\delta(h,\alpha)$ and 
$\gamma_{\Gamma}(h,\alpha)$ can be expressed in terms of the 
coefficients in front of  ${\varepsilon}^{-1}$
in the decompositions of the corresponding renormalization constants $Z$
\cite{'tHooft:1973mm},
\cite{Collins:1973yy}:
\begin{eqnarray}
& &\beta(h)=\left( h {\partial \over \partial h}-1 \right)a^{(1)}(h)
\nonumber \\
& &~~~~~~=h[2\widetilde{\gamma}_1(h,\alpha)-\gamma_3(h,\alpha)
-2\widetilde{\gamma}_3(h,\alpha)]
=-\beta_1h^2-\beta_2h^3-\beta_3h^4-\cdots,
\nonumber 
\\
& &\gamma_m(h)=h{\partial b^{(1)}(h) \over \partial h}=
-\gamma_1h-\gamma_2h^2-\gamma_3h^3-\cdots,
\nonumber 
\\
& &\delta (h,\alpha)=h{\partial d^{(1)}(h) \over \partial h}=
\gamma_3(h,\alpha),
\nonumber
\\
& &\gamma_{\Gamma}(h,\alpha)=h{\partial c^{(1)}(h) \over \partial h},
\label{RGfunctions}
\end{eqnarray}
where $\widetilde{\gamma}_1(h)$ is the anomalous dimension of the ghost-ghost-gluon
vertex, $\gamma_3$ and $\widetilde{\gamma}_3$ are  anomalous dimensions
of the gluon and ghost propagators respectively. The coefficients in
front of   $1/\varepsilon^{\nu} (\nu\geq 2)$  for the renoramlization 
constants $Z_{\Gamma}$ and $Z_m$  are related to $c^{(1)}$ and $b^{(1)}$ 
by  \cite{'tHooft:1973mm}, \cite{Collins:1973yy}:
\begin{eqnarray*}
& & \left\{\beta(h){\partial \over \partial h}
+\gamma_3(h,\alpha)
\alpha {\partial \over \partial \alpha}+\gamma_\Gamma (h,\alpha)\right\}
c^{(\nu)}(h,\alpha)=h{\partial \over \partial h}c^{(\nu+1)}(h,\alpha), \\
& &\left\{\gamma_m(h)+\beta(h){\partial \over \partial h}\right\}
b^{(\nu)}(h)=h{\partial \over \partial h}b^{(\nu+1)}(h).
\end{eqnarray*}
These relationships were used to partially verify our calculations.

The  goal of this paper is to calculate the three-loop
approximation for  $\gamma_m(h)$. The two-loop calculations were presented 
in Refs.\cite{Tarrach:1980up}.
 To find  $\gamma_m(h)$, it is convenient to use the relation
$\gamma_m(h)=-\gamma_2(h,\alpha)+\gamma_{\overline{\psi} \psi}(h,\alpha)$ that
follows from the equation
\begin{eqnarray*}
Z_m=Z_{\overline{\psi} \psi}Z^{-1}_2
\end{eqnarray*}
where
$$
\gamma_{\overline{\psi} \psi}(h,\alpha)=-\left. \mu^2{\partial \over \partial \mu^2}
\ln Z_{\overline{\psi} \psi}\right|_{h_B,m_B,\alpha_B-fixed},
$$
and $Z_{\overline{\psi} \psi}$ is the renormalization constant of the two-point 
fermion Green function with the insertion $\overline \psi(y) \psi (y)$, i.e
$$
\langle \overline{\psi}(x) \psi (0) \int d^ny \overline \psi(y) \psi(y) \rangle.
$$

According to the method described in \cite{Vladimirov:1977ak},
 for the renormalization constants $Z_{\overline{\psi} \psi}$ and
$Z_2$ we shall use	the following representation:
\begin{eqnarray*}
& &Z_{\overline{\psi} \psi}=1-KR'\Gamma_{\overline{\psi} \psi},\\
& &Z^{-1}_2=1-KR'\Gamma_2.
\end{eqnarray*}
Here $K$ is an operator that removes regular in $\varepsilon$ terms
$$
K\sum_{\nu}b_{\nu}\varepsilon^{\nu}=\sum_{\nu<0} b_{\nu}\varepsilon^{\nu},
$$
and $R'$ is the  $R$ operation for the minimal subtraction scheme without 
the last subtraction applied to the Green function $\Gamma$, i.e
$$
R=\left(1-K\right)R'.
$$

We apply $K$ and $R'$ operations to the  diagrams of the following type:\\
\begin{figure}[h]
\begin{center}
\includegraphics[scale=1.1]{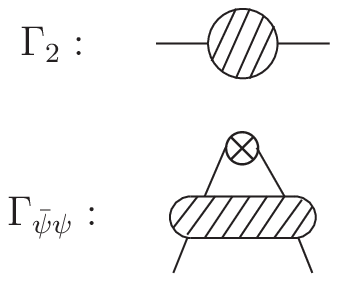}
\end{center}
\end{figure}

\noindent
where $-\!\!\bigotimes\!\!-$ corresponds to the insertion $\int \overline
\psi(y) \psi(y)d^ny$.
\\

As an illustrative example we consider calculation of the contribution to
$Z_{\overline \psi \psi}$ from one of the three-loop diagrams. For example,
\begin{figure}[h]
\begin{center}
\includegraphics[scale=1.1]{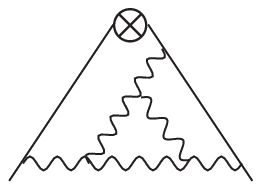}
\end{center}
\end{figure}
\vspace{-0.4cm}

\noindent
The $R'$ for this diagram with  the combinatorial factor and appropriate counterterms can
schematically  be represented as:
\begin{figure}[h]
\begin{center}
\includegraphics[scale=1.0]{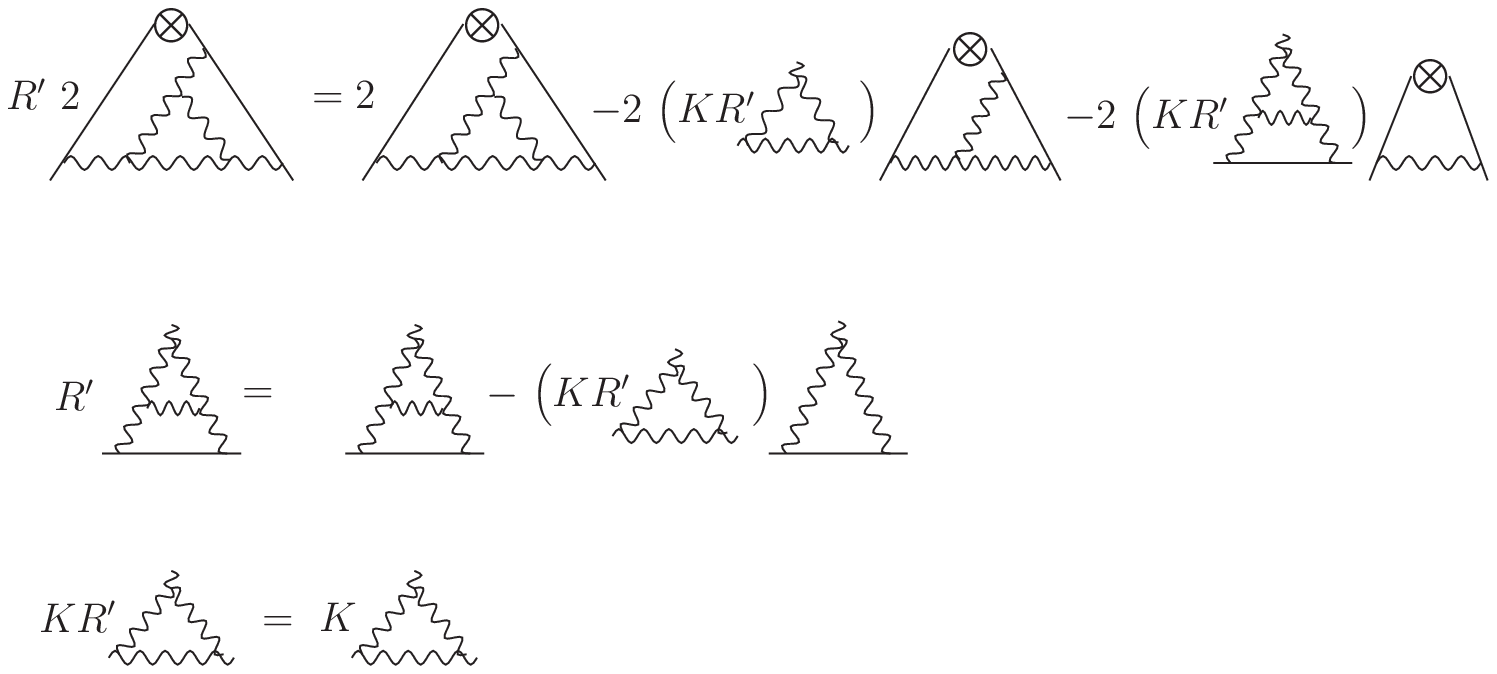}
\end{center}
\end{figure}
\vspace{-0.4cm}

\noindent
\vspace{0.3cm}
\noindent
Substituting  expressions for the one- and two- loop diagrams and multiplying 
them by appropriate  $KR'$,  we get for the diagram, 
counterterms with two- and one- loop subdiagrams
the following expression:
\\
\begin{figure}[h]
\begin{center}
\includegraphics[scale=1.0]{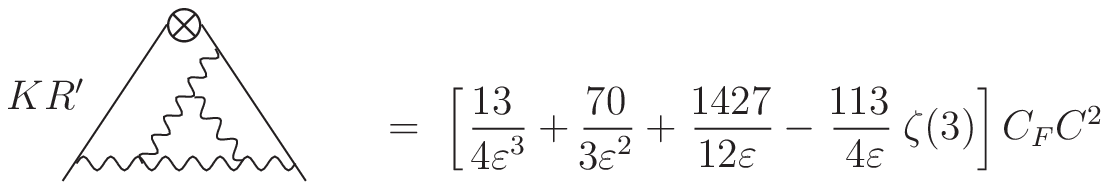}
\end{center}
\vspace{-1cm}
\end{figure}
\vspace{-3.0cm}
\begin{eqnarray*}
&~~~~~~~~~~~~~~~~~~~~~~~~~~+&\!\left[\frac{39}{4\varepsilon^3}+\frac{19}{4\varepsilon^2}+
\frac{231}{16\varepsilon} \right]C_FC^2  \nonumber \\
&~~~~~~~~~~~~~~~~~~~~~~~~~~-&\!\left[\frac{39}{4\varepsilon^3}+\frac{299}{8\varepsilon^2}+
\frac{975}{8\varepsilon}
-\frac{117}{4\varepsilon}\zeta(3)\right]C_FC^2  \nonumber \\
&~~~~~~~~~~~~~~~~~~~~~~~~~~=&\!\left[\frac{13}{4\varepsilon^3}-\frac{223}{24\varepsilon^2}+
\frac{551}{48\varepsilon}
+\frac{1}{\varepsilon}\zeta(3)\right]C_FC^2.
\end{eqnarray*}

All calculations were performed on the CDC-6500 computer using the algebraic
manipulation system SCHOONSCHIP \cite{Strubbe:1974vj}.
In Appendix, we present contributions to $Z^{-1}_2$ and $Z_{\overline{\psi} \psi}$
from diagrams with one- and two- loop insertions, factorized diagrams
and 1PI diagrams.
At the three loop level we get:
\begin{eqnarray*}
Z^{-1}_2&=& 1+\frac1{\varepsilon}hC_F+\left[\frac1{\varepsilon^2} \left(
\frac12
C_F-C \right)+\frac1{\varepsilon}\left(\frac{17}4C-tf-\frac34C_F\right)
\right]h^2C_F \\
& & +\left[\frac1{\varepsilon^3}\left(\frac74C^2-\frac13Ctf-CC_F+
 \frac16C^2_F \right) \right. \\
& & -\frac1{\varepsilon^2} \left( \frac{104}9C^2-\frac{56}9Ctf+\frac89t^2f^2-
\frac{73}{12} CC_F+\frac53C_Ftf+\frac34C^2_F \right) \\
& & -\frac1{\varepsilon} \left(  \frac{1301}{108}Ctf+\frac{143}{12}CC_F
-\frac{10559}{432}C^2-\frac{20}{27}t^2f^2 \right.\\
& &~~~~~~~~~~~~~~~~~~~~\left. \left.-C_Ftf-\frac12C^2_F-4CC_F\zeta(3)
+\frac52C^2\zeta(3) \right)  \right]h^3C_F,
\end{eqnarray*}

\begin{eqnarray*}
\lefteqn{Z_m=1-3C_Fh\frac1{\varepsilon}+\left[\frac1{\varepsilon^2}
\left(\frac{11}2C-2tf+\frac92C_F\right)+\frac1{\varepsilon}
\left(-\frac{97}{12}C+\frac53tf-\frac34C_F\right)\right]h^2C_F}\\
 & &+\left[\frac1{\varepsilon^3}\left(\frac{88}9Ctf-\frac{33}2CC_F-\frac{121}9C^2-
\frac{16}9t^2f^2+6C_Ftf-\frac92C^2_F\right) \right. \\
 & &+\frac1{\varepsilon^2}\left(-\frac{484}{27}Ctf+\frac{313}{12}CC_F
+\frac{1679}{54}C^2+\frac{40}{27}t^2f^2-\frac{29}3C_Ftf+\frac94C^2_F\right)\\
 & &+\frac1{\varepsilon}\left(\frac{556}{81}Ctf+
\frac{43}4CC_F-\frac{11413}{324}C^2
 +\frac{140}{81}t^2f^2+\frac{46}3C_Ftf \right.    \\
& &~~~~~~~~~~~~~~~~~~~~\left. \left.-\frac{43}2C^2_F+16\zeta(3)tf(C-C_F)\right)\right]h^3C_F.
\end{eqnarray*}

Using (\ref{massaB}) and (\ref{RGfunctions}) we get
\begin{eqnarray*}
&&\gamma_m(h)=-3C_Fh-\left(\frac{97}6C-\frac{10}3tf+
\frac32C_F\right)C_Fh^2+\left(\frac{556}{27}Ctf+\frac{129}4CC_F
\right. \nonumber \\
&&~-\left. \frac{11413}{108}C^2
 +\frac{140}{27}t^2f^2+46C_Ftf -\frac{129}2C^2_F+
48\zeta(3)tf(C-C_F)\right)h^3C_F,
\end{eqnarray*}
\begin{eqnarray}
&&\gamma_2(h,1)=-C_Fh-\left(\frac{17}2C-2tf-\frac32C_F\right)C_Fh^2-
\left( \frac{10559}{144}
C^2- \frac{1301}{36}Ctf \right. \nonumber \\
&&~~\left.
-\frac{143}{4}CC_F+\frac{20}{9}t^2f^2+3C_Ftf+\frac32C^2_F+12CC_F\zeta(3)
-\frac{15}2C^2\zeta(3)  \right) h^3C_F.
\label{gammaM}
\end{eqnarray}
At the two loop level our results coincide with that presented in 
\cite{Tarrach:1980up}.
Unlike the  $\beta$- function, the three-loop contribution to $\gamma_m$
depens on the  Riemann function \(\zeta(3)=\sum_{l=1}^{\infty}1/l^3 \).
The coefficient in $\gamma_m$ proportional to $C_Ft^2f^2$ coincides with that
predicted in \cite{Espriu:1982pb}.
 For the quantum chromodynamics ($SU(3)$ gauge group with
$C=3$, $C_F=4/3$, $t=1/2$) we have:
\begin{eqnarray*}
\gamma_{m_{i}}(h)=-4h-\left(\frac{202}3- \frac{20}9f \right)h^2-
\left( \frac{3747}3 -\frac{2216}{27}f-\frac{140}{81}f^2
\right)h^3 
+\frac{160}3\zeta(3)f h^3,
\end{eqnarray*}
\begin{eqnarray*}
\gamma_2(h,1)=-\frac43h-\left(\frac{97}3+\frac43f\right)h^2
-\left( \frac{24941}{36} - \frac{1253}{18}f + \frac{20}{27}f^2-26\zeta(3)  \right)h^3.
\end{eqnarray*}
For completeness we present also expression for the $\beta$- function 
\cite{Tarasov:2013zv}:
\begin{eqnarray*}
\beta(h)=-\left(11-\frac23f\right)h^2-\left(102-\frac{38}3f\right)h^3
-\left(\frac{2857}2-\frac{5033}{18}f+\frac{325}{54}f^2\right)h^4.
\end{eqnarray*}
From (\ref{gammaM}),  one can easily get $\gamma_2$ and $\gamma_m$ 
for the quantum
electrodynamics, by setting $C=0$, $C_F=1$, $tf=1$:
\begin{eqnarray*}
\gamma_{m}(h)&=&-3h+\frac{11}6h^2-\frac{719}{54}h^3-48\zeta(3)h^3,\\
\gamma_2(h) &=&-h+\frac72h^2-\frac{121}{18}h^3.
\end{eqnarray*}

The solutions of the renormalization group equation (2) in the t'Hooft 
scheme are expressed in terms of the effective parameters $\overline h$, $\overline m$,
$\overline{\alpha}$ determined by the system of equations:
\begin{eqnarray*}
& &{d \overline{h} \over dl}=\beta(\overline{h}), ~~~~~~~~~~~~~~~
l=\ln \frac{Q^2}{\mu^2},\\
& &{d \ln \overline{\alpha} \over dl}=\delta(\overline h,\overline{\alpha}), \\
& &{d \ln \overline{m}_i \over dl}=\gamma_{m_{i}}(\overline{h}).
\end{eqnarray*}
The 
expression for $\overline{h}(L)$ up to three loops was given in
\cite{Tarasov:2013zv}:

\begin{eqnarray*}
\overline{h}(L)&=& \frac1{\beta_1L}- \frac{\beta_2}{\beta_1^3} \frac{\ln L}{L^2}+
{\beta^2_2(\ln^2L-\ln L) \over \beta_1^5 L^3}+{\beta_3\beta_1-\beta_2^2
\over \beta_1^5L^3} \nonumber\\
&-& {\beta_2^3 \ln^3L \over \beta_1^7 L^4}+{5\beta_2^3 \ln^2L \over
    2\beta_1^7 L^4}
+ {(-3\beta_1\beta_2\beta_3+2\beta_2^3) \ln L \over \beta_1^7 L^4}
+O \left( \frac1{L^4} \right)
\end{eqnarray*}
Here $L=\ln \frac{Q^2}{\Lambda^2}\equiv\int\frac{dx}{\beta(x)}+\ln \frac{Q^2}
{\mu^2}$. An arbitrary integration constant is chosen   to cancel the
term with $1/L^2$ in the  decomposition of $\overline{h} (L)$.

The solution for $\overline{m}_{i}$ is:
\begin{eqnarray*}
&&\overline{m}_i(L) = {\hat{m}_i(L)}^{-\frac{\gamma_1}{\beta_1}}
 \left\{1  +
     \frac{\left(\gamma_2 \beta_1-\beta_2\gamma_1\right)}{\beta_1^3 L} 
       -\frac{ \beta_2 \gamma_1}{\beta_1^3} \frac{\ln L }{L}
\right.
\\
 &&
     +\frac{1}{2 \beta_1^6}\Bigl(
       \beta_1^3 \gamma_3 + \beta_1^2 \gamma_2^2-\beta_1^2 \beta_2
       \gamma_2+\beta_1^2 \beta_3 \gamma_1
        -\beta_1 \beta_2^2 \gamma_1-2 \beta_1 \beta_2 \gamma_1
     \gamma_2+\beta_2^2 \gamma_1^2 \Bigr)
\frac{1}{L^2}
\\
 && 
\left. 
 +\frac{1}{\beta_1^6}
    \Bigl(\beta_2^2 \gamma_1^2-\beta_1^2 \beta_2 \gamma_2-\beta_1 \beta_2 \gamma_1
    \gamma_2 \Bigr)
     \frac{\ln L}{ L^2}
+\frac{1}{2 \beta_1^6} 
      \left(\beta_1 \beta_2^2
  \gamma_1+\beta_2^2\gamma_1^2\right)\frac{ \ln^2 L}{L^2}
  \right\}.
\end{eqnarray*}

At the present time  $\delta(h,\alpha)$  in an arbitrary gauge at the three-loop level
is not known.  But in the MS scheme physical quantities depend
only on $\overline{h}$, so for them we can use the renormalization group at the
three-loop level.

The most interesting application of our results can be the $m_b/ m_\tau$
ratio in the $SU(5)$ grand unified theory and calculation of $\gamma_m(h)$
at three loop level in supersymmetric theories
\cite{Tarasov:2013zv}.

\underbar{Acknowledgments}

I am very grateful to  D.V.Shirkov,  A.L. Kataev and
 D.I.Kazakov for valuable discussions.

The present text  was
published in 1982 as a JINR Communication, JINR-P2-82-900.

\section{Appendix}
The three-loop 1PI diagrams  of the fermion propagator.
\begin{figure}[h]
\begin{center}
\includegraphics[scale=1.0]{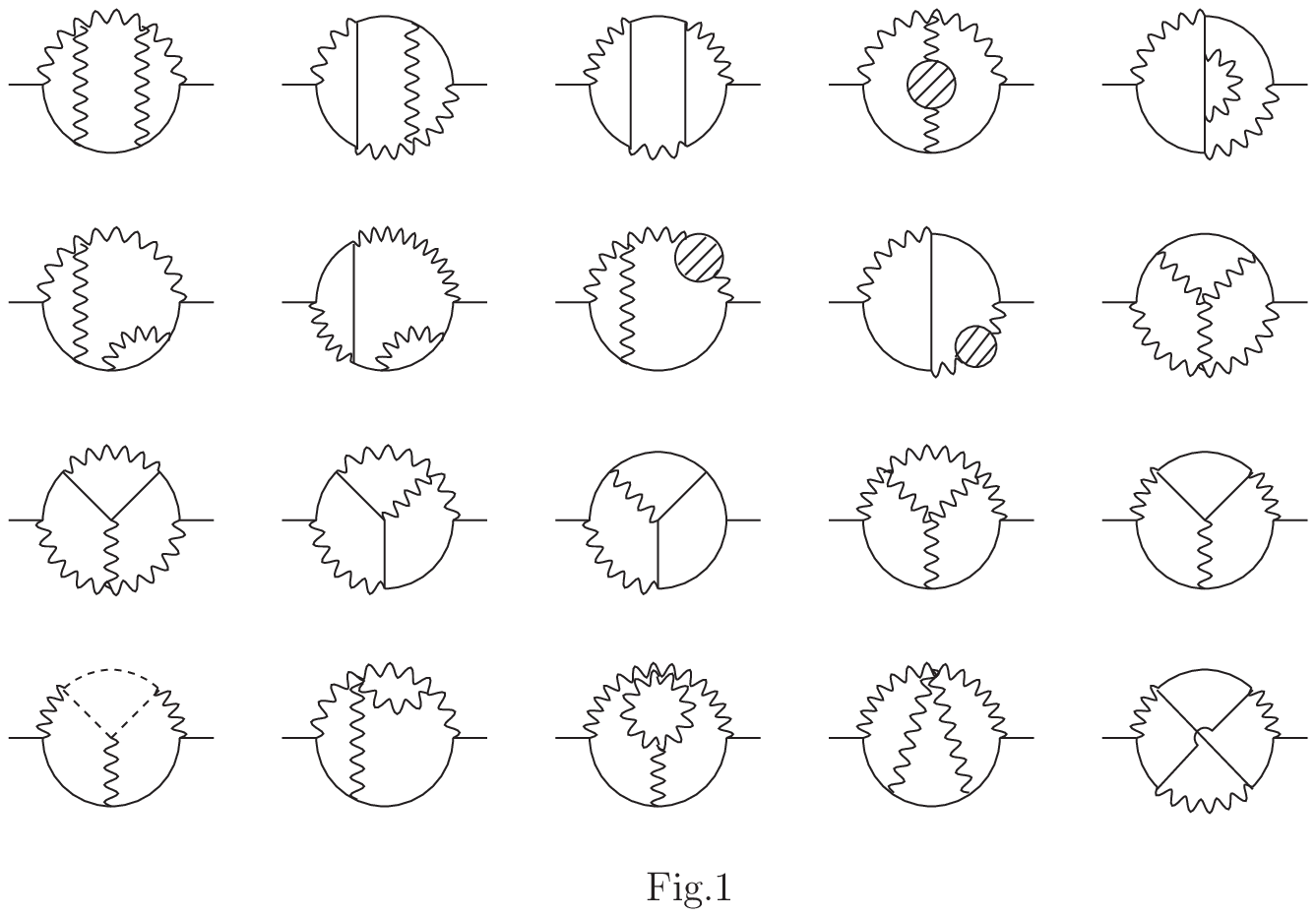}
\end{center}
\end{figure}
\begin{figure}[h]
\begin{center}
\includegraphics[scale=1.0]{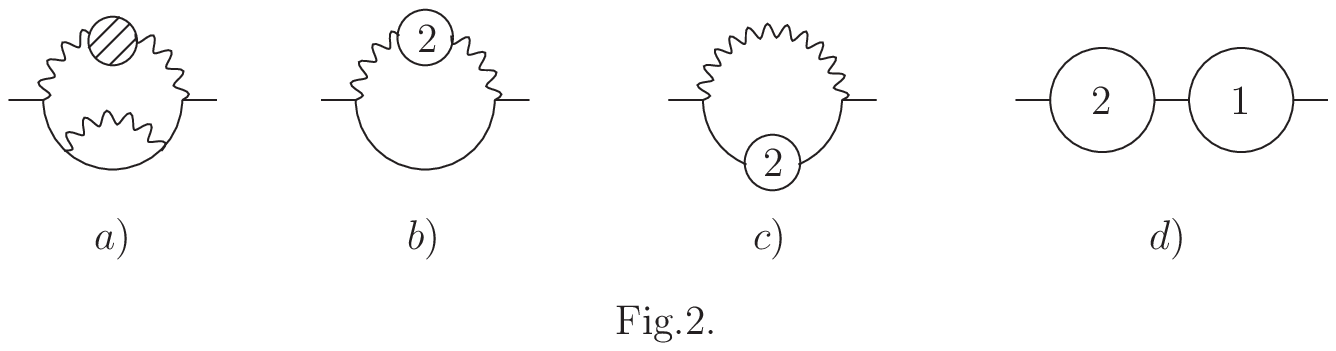}
\end{center}
\end{figure}

\newpage

The contributions to $Z_2^{-1}$ from different sets of the  diagrams ($KR'$):\\
Fig.1.
\begin{eqnarray*}
& &-\frac{C_F}{\varepsilon^3}\left[-\frac74C^2-\frac43CC_F+\frac13Ctf\right]\\
& &-\frac{C_F}{\varepsilon^2}\left[\frac{219}{24}C^2+\frac{35}6CC_F-\frac43C_F
-\frac{19}6Ctf-\frac43C_Ftf\right]\\
& &-\frac{C_F}{\varepsilon}\left[-\frac{233}{12}C^2+\frac{17}2CC_F-\frac7{12}
C^2_F+\frac{43}6Ctf-\frac73C_Ftf\right]
-\frac{C_F}{\varepsilon}C\zeta(3)\left(\frac52C-2C_F\right).
\end{eqnarray*}
Fig.2a:
$$
\frac{C_F^2}{\varepsilon^2}\left(\frac5{12}C-\frac13tf\right)
+\frac{C_F^2}{\varepsilon}\left(\frac38C-\frac16tf\right).
$$
Fig.2b:
\begin{eqnarray*}
- \frac{C_F}{\varepsilon^2} \Bigl[\frac{175}{72}C^2-\frac{55}{18}Ctf
+ \frac89t^2f^2  \Bigr] 
- \frac{C_F}{\varepsilon} \left[-\frac{2171}{432}C^2+\frac{527}{108}Ctf+
2C_Ftf-\frac{20}{27}t^2f^2 \right] .
\end{eqnarray*}
Fig.2c:
\begin{eqnarray*}
-\frac{C_F^2}{\varepsilon^3} \left[ \frac13C-\frac16C_F\right]
+\frac{C_F^2}
{\varepsilon^2} \Bigl[3C-\frac7{12}C_F &-& \frac23tf
 \Bigr]\\
&-& \frac{C_F^2}{\varepsilon}\left[\frac{91}{24}C-2\zeta(3)C+\frac1{12}C_F-
\frac56tf\right].
\end{eqnarray*}
Fig.2d: $~~~~~~~~-\frac2{\varepsilon^3}C_F^2C$
\\
The contributions to $Z_{\overline{\psi}\psi}$  from the diagrams with the
 insertion
$-\!\!\bigotimes\!\!-$  were calculated simultaneously with the diagrams
shown in Fig.1 and Fig.2. In fact, for the fermion propagator instead of
$\hat{p}/p^2$ we used $(\hat{p}+m)/p^2$, and after multiplication of all
propagators we kept only the part without the mass term and the
 part linear in $m$.
The coefficient in front of  $m$  determines the contribution of the diagrams with the mass insertion. For example, from the diagram
\begin{figure}[h]
\begin{center}
\includegraphics[scale=1.0]{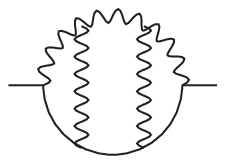}
\end{center}
\end{figure}
\noindent
\newpage
we get the following diagrams with the insertion:
\begin{figure}[h]
\begin{center}
\includegraphics[scale=1.0]{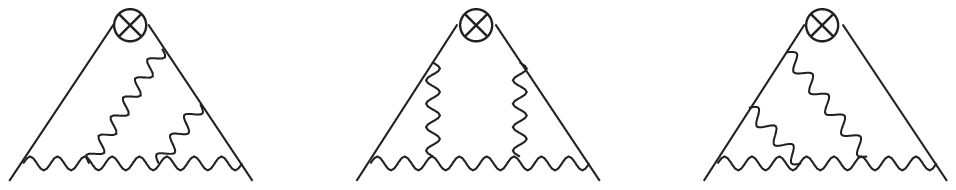}
\end{center}
\end{figure}

The
contributions to $Z_{\overline{\psi} \psi}$ from different sets of diagrams ($KR'$)\\

\vspace{0.5cm}
Fig.1:
\vspace{-0.3cm}
\begin{eqnarray*}
&-&\frac{C_F}{\varepsilon^3} \left[\frac{31}3C^2+\frac{50}3CC_F+8C_F^2-4Ctf-
\frac83C_Ftf \right] \\
&-&\frac{C_F}{\varepsilon^2}\left[-\frac{181}6C^2-\frac{80}3C_FC+8C_F^2
+\frac{34}3Ctf+\frac83C_Ftf\right]\\
&-&\frac{C_F}{\varepsilon}\left[\frac{181}4C^2-\frac{145}3CC_F+17C_F^2
-13Ctf+\frac{20}3C_Ftf\right]\\
&-&\frac{C_F}{\varepsilon}\zeta(3)\left[-\frac32C^2+20C_FC-16C_F^2-8Ctf\right].
\end{eqnarray*}

Fig.2a:
$$
-\frac{C_F^2}{\varepsilon^3}\left[\frac{10}3-\frac83tf\right]-\frac{C_F^2}
{\varepsilon^2}\left[-\frac{17}3C+4tf\right]-\frac{C_F^2}{\varepsilon}
\left[-C+\frac43tf \right].
$$

Fig.2b:
\begin{eqnarray*}
&-&\frac{C_F}{\varepsilon^3}\left[\frac{175}{36}C^2-\frac{55}9Ctf+
\frac{16}9t^2f^2\right] \\
&-&\frac{C_F}{\varepsilon^2}\left[-\frac{337}{27}C^2+\frac{346}{27}Ctf+
4C_Ftf-\frac{64}{27}t^2f^2\right]\\
&-&\frac{C_F}{\varepsilon}\left[\frac{18685}{1296}C^2-\frac{1915}{324}Ctf
-17C_Ftf-\frac{80}{81}t^2f^2\right]\\
&-&\frac{C_F}{\varepsilon}\zeta(3)\left[-C^2-8Ctf+16C_Ftf\right]
\end{eqnarray*}

Fig.2c:
\begin{eqnarray*}
&-&\frac{C_F^2}{\varepsilon^3}\left[6C+\frac83C_F-\frac83tf\right]
-\frac{C_F^2}{\varepsilon^2}\left[-17C-10C_F+4tf\right]\\
&-&\frac{C_F^2}{\varepsilon}\left[\frac{80}3C+5C_F-\frac{16}3tf
-16\zeta(3)C+16\zeta(3)C_F\right]
\end{eqnarray*}

Fig.2d: 0.

\end{document}